\documentclass[twocolumn,showpacs,preprintnumbers,amsmath,amssymb]{revtex4}

\usepackage{graphicx}
\usepackage{dcolumn}
\usepackage{bm}

\begin{document}

\preprint{APS/123-QED}

\title{Condensation phenomena with distinguishable particles}

\author{Jun Ohkubo}
\email[Email address: ]{ohkubo@issp.u-tokyo.ac.jp}
\affiliation{
Institute for Solid State Physics, University of Tokyo, 
Kashiwanoha 5-1-5, Kashiwa-shi, Chiba 277-8581, Japan
}
\date{\today}

\begin{abstract}
We study real-space condensation phenomena in a type of classical stochastic processes (site-particle system),
such as zero-range processes and urn models.
We here study a stochastic process in the Ehrenfest class,
i.e., particles in a site are distinguishable.
In terms of the statistical mechanical analogue, the Ehrenfest class obeys the Maxwell-Boltzmann statistics.
We analytically clarify conditions for condensation phenomena in disordered cases in the Ehrenfest class.
In addition, we discuss the preferential urn model as an example of the disordered urn model.
It becomes clear that the quenched disorder property plays an important role
in the occurrence of the condensation phenomenon in the preferential urn model.
It is revealed that the preferential urn model shows three types of condensation depending on the disorder parameters.
\end{abstract}

\pacs{75.10.Nr, 02.50.Ey, 64.60.-i, 05.40.-a}
\maketitle

\section{Introduction}

Condensation phenomena have been widely observed in various stochastic processes and physical contexts,
such as mass transport models \cite{Majumdar2005,Evans2006,Evans2006a},
traffic flow \cite{Evans1996}, condensation of edges in complex networks \cite{Bianconi2001}.
It is notable that some of the condensation phenomena occur in nonequilibrium systems,
such as asymmetric simple exclusion process (ASEP) and zero-range process (ZRP) \cite{Evans2005}.
In general, it is difficult to study such nonequilibrium systems,
but the steady state properties in a class of the stochastic processes can be treated analytically
because their partition functions are written in simple factorized forms.
Such a simple form of the partition function enables us to study the condensation phenomenon analytically.

Mass transport models, such as the ZRP and urn models 
\cite{Ritort1995,Bialas1997,Drouffe1998,Bialas1998,Bialas2000,Godreche2001,Leuzzi2002},
have been widely used in order to research characteristic features of the condensation phenomena.
When the condensation arises, a single site contains a macroscopic number of particles.
Both the ZRP and urn models consist of many sites containing an integer number of particles.
Inter-particle interactions are modeled by allowing the particle transition rate to depend on
the number of particles in each site.
Furthermore, in order to discuss general situations,
it is possible to consider the case in which the transition rate depends also on a site-dependent disorder parameter.
The mathematical structures for the ZRP and the urn models are almost the same,
so that one can treat these models in a similar manner.

In the urn models, there are two types of statistics \cite{Godreche2001}:
one is the Ehrenfest class in which particles in a site are distinguishable;
the other is the Monkey class which consists of indistinguishable particles.
It has been revealed that the Ehrenfest class corresponds to the statistical mechanics with
the Maxwell-Boltzmann statistics.
Monkey class is related to the Bose-Einstein statistics, and in general raises real-space condensation phenomena
\cite{Bialas1997,Godreche2001}.
The real-space condensation phenomenon means that a site has a macroscopic number of particles.

We here note that some of recent researches on complex networks have dealt with 
distinguishable components \cite{Albert2000,Ohkubo2005a,Ohkubo2005b,Ohkubo2005},
and a type of condensation phenomena has been observed numerically 
in a network model in the canonical ensemble, in which the total number of nodes and edges do not change
\cite{Ohkubo2005b}.
When we map a complex network to an urn model,
the distinguishable property of edges in the network model corresponds to 
that of particles in the urn model \cite{Ohkubo2005}.
We therefore expect that the condensation phenomenon is able to occur even in the
Ehrenfest class (Maxwell-Boltzmann statistics)
due to the occurrence of the condensation phenomenon in the complex networks.
Actually, there are a few example of the Ehrenfest class, 
in which the real-space condensation phenomenon can arise \cite{Lipowski2002,Torok2005}.
However, in the complex networks, 
quenched disorder in the model is expected to be important,
and discussion between the quenched disorder and the condensation phenomenon in the Ehrenfest class
has not been given yet.

In the present paper, we consider stochastic processes 
with the Maxwell-Boltzmann statistics, i.e., the Ehrenfest class, and with the quenched disorder.
Conditions for the occurrence of the condensation phenomenon are clarified.
Furthermore, we study one example for the condensation phenomenon in the Ehrenfest class with the
quenched disorder, i.e., the preferential urn model proposed in Ref.~\cite{Ohkubo2005}.
We analytically and numerically confirm the condensation phenomenon in the preferential urn model.
It is revealed that there are three types of condensation phenomena depending on the disorder parameters.

The construction of the present paper is as follows.
In Sec.~II, we will present examples of the stochastic models in the Ehrenfest class (the Maxwell-Boltzmann statistics)
and explain their partition functions.
The conditions for the condensation phenomenon in disordered cases are explained in Sec.~III.
In Sec.~IV, we give a concrete example for the condensation, i.e., the preferential urn model,
and discuss three types of the condensation phenomena.
We also give some numerical results in order to confirm the condensation phenomena in Sec.~IV.
Section~V gives concluding remarks.

\section{Models in the Ehrenfest class}

\begin{table*}
\caption{Comparison between the Ehrenfest class and the Monkey class.}
\label{table}
\begin{tabular}{lll}
\hline \hline
  &  Ehrenfest class $\qquad$ & Monkey class \\ \hline
Particles in a site  &  distinguishable  &  indistinguishable\\
Weight of site $i$ &  $p(h_i,n_i)/n_i!$ & $p(h_i,n_i)$ \\
Statistical-mechanical analogue $\qquad$ &  Maxwell-Boltzmann  & Bose-Einstein \\
Condensation phenomenon &  difficult to occur & easy to occur \\
\hline \hline
\end{tabular}
\end{table*}

\subsection{Examples of models}

We study stochastic processes with a factorized partition functions.
Examples of these stochastic processes are the ZRP and the urn models, as described in Sec.~I.
There are two types of statistics \cite{Godreche2001}: One is the Ehrenfest class;
the other is the Monkey class.
The characteristics of these two classes are summarized in Table.~\ref{table}.
The Ehrenfest class has distinguishable particles, so that it corresponds to the Maxwell-Boltzmann statistics.
In contrast, indistinguishable particles are considered in the Monkey class.
Here, we give two examples of the stochastic processes in the Ehrenfest class and with quenched disorder.

\begin{figure}
\begin{center}
  \includegraphics[width=80mm,keepaspectratio,clip]{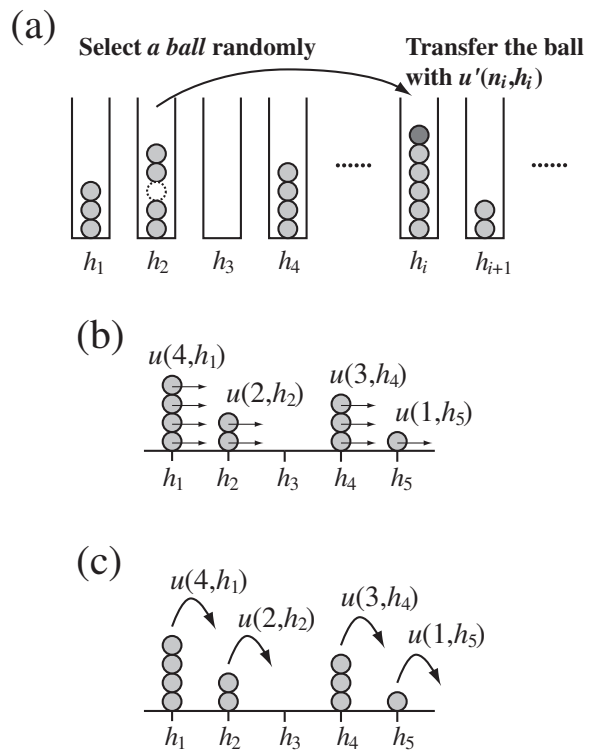} 
\caption{
(a) The urn model in the Ehrenfest class (the Maxwell-Boltzmann statistics).
(b) ZRP in the Ehrenfest class, in which each particle can hop with rate $u(h,n)$.
(c) The most general ZRP (the Monkey class), in which only one particle in a site hops out with rate $u(h,n)$. 
}
\label{fig_model}
\end{center}
\end{figure}

\begin{itemize}
\item
\textit{Urn model on a complete graph in Ehrenfest class}:
The urn model on a complete graph consists of $L$ sites (urns) and $N$ particles.
The density $\rho$ is defined by $\rho \equiv N / L$.
An disorder parameter $h_i$ is assigned to each site by a probability density $\phi(h)$.
The number of particles in site $i$ is denoted by $n_i$.
When we adopt a Monte Carlo method (the heat-bath rule) for the dynamics,
a particle is selected randomly in one Monte Carlo step,
and then transfered to an urn selected by a probability $u'(h_i,n_i)$ (Fig.~\ref{fig_model}(a)) 
\cite{Godreche2001}.
Note that the probability depends on the number of particles and the disorder parameter of the arrival site;
it does not depend on the properties at the departure site.

\item
\textit{ZRP with the Maxwell-Boltzmann statistics in one-dimensional lattice with periodic boundary conditions}:
The lattice consists of $L$ sites, and there are $N$ particles.
Each particle hops out to the nearest neighboring site, only in one direction.
A particle is selected by a probability $u(h_i,n_i)$ in one Monte Carlo step,
and hops out (Fig.~\ref{fig_model}(b)).
\end{itemize}

We note that the most general ZRP has not the Maxwell-Boltzmann characteristics.
The dynamics of the most general ZRP is as follows; ``a site is selected by a probability $u(h_i,n_i)$,
and a particle in the selected site hops out (Fig.~\ref{fig_model}(c)).'' 
Hence, the particles in a site are indistinguishable,
so that the most general ZRP corresponds to the Monkey class.

In Sec.~IV, 
we will specify an urn model in the Ehrenfest class in more detail
in order to investigate the condensation phenomenon in disordered cases with the Maxwell-Boltzmann statistics.

\subsection{Partition function}

For the Monkey class (e.g., the most general ZRP), the partition function is written as \cite{Godreche2001}
\begin{align}
Z_{L,N}^\textrm{Monkey} &= 
\sum_{n_1 = 0}^\infty \dots \sum_{n_L = 0}^\infty p(h_1,n_1) \dots p(h_L,n_L) \notag \\
& \times \delta \left( \sum_{i=1}^{L} n_i, N \right), 
\end{align}
where $\delta \left( \sum_{i=1}^{N} n_i, M \right)$ is Kronecker delta,
which indicates that the total number of particles is conserved.
The weight $p(h_i,n_i)$ is calculated from the transition rate of the ZRP, $u(h_i,n_i)$, by 
$p(h_i,n_i) = 1 / \prod_{j=1}^{n_i} u(h_i,j)$.
For the urn models, we calculate the weight $p(h_i,n_i)$ as $p(h_i,n_i) = \prod_{j=1}^{n_i} u'(h_i,j)$.

The partition function of the Ehrenfest class (Maxwell-Boltzmann statistics) is a similar one with 
that of the Monkey class;
\begin{align}
Z_{L,N} &= 
\sum_{n_1 = 0}^\infty \dots \sum_{n_L = 0}^\infty \frac{p(h_1,n_1)}{n_1!} \dots \frac{p(h_L,n_L)}{n_L!} \notag \\
& \times  \delta \left( \sum_{i=1}^{L} n_i, N \right).
\label{eq_partition_function}
\end{align}
Note that the partition function involves factorials $\{1/(n_i!)\}$,
which stems from the distinguishable property of particles \cite{Godreche2001}.
The factor $1/(n_i!)$ plays an important role for the condensation phenomenon,
which will be explained in the subsequent section.

\section{Condensation in disordered cases}

From the partition function of Eq.~\eqref{eq_partition_function},
the steady-state occupation distribution is derived.
Since we consider site-dependent disordered cases,
we analyze it with the replica method, as described in Ref.~\cite{Ohkubo2006}.
We here briefly summarize the analysis \cite{note_replica}.

Using the partition function of Eq.~\eqref{eq_partition_function},
the probability with which site (urn) $1$ has $k$ particles is given by
\begin{align}
&\frac{1}{Z_{L,N}}  
\sum_{n_1 = 0}^\infty \dots \sum_{n_L=0}^\infty \delta(n_1,k) \frac{p(h_1,n_1)}{n_1!} \dots \frac{p(h_L,n_L)}{n_L!} 
\notag \\
& \quad \times \delta\left( \sum_{i=1}^{L} n_i , N\right) \notag \\
&= \frac{p(h_1,k)}{k!} \frac{Z_{L,N}'}{Z_{L,N}},
\end{align}
where we define $Z_{L,N}'$ as
\begin{align}
Z_{L,N}' =& 
\sum_{n_2 = 0}^\infty \dots \sum_{n_L = 0}^\infty
\frac{p(h_2,n_2)}{n_2!} \dots \frac{p(h_L,n_L)}{n_L!}  \notag \\
& \times \delta\left( \sum_{i=2}^{L} n_i , N-k\right)
\end{align}
for notational simplicity.
We therefore obtain the occupation distribution as
\begin{align}
P(k) =& 
\left\langle 
\frac{p(h_1,k)}{k!}
\frac{Z_{L,N}'}{Z_{L,N}}
\right\rangle_{\{ h_1, \dots, h_L\}} \notag \\
=& \int dh_1 \,\, \phi(h_1) \frac{p(h_1,k)}{k!}
\left\langle 
\frac{Z_{L,N}'}{Z_{L,N}}
\right\rangle_{\{ h_2, \dots, h_L\}}  \notag \\
=& \int dh_1 \,\, \phi(h_1) \frac{p(h_1,k)}{k!} \notag \\
& \times \exp\left[ 
- \left\langle \ln Z_{L,N} \right\rangle_{\{ h_2, \dots, h_L\}} 
+ \left\langle \ln Z_{L,N}' \right\rangle_{\{ h_2, \dots, h_L\}} 
\right]
\label{eq_pre_occ}
\end{align}
where $\langle \cdots \rangle_{\{ h_1, h_2, \dots, h_L\}}$ is the configurational average,
and $\phi(h)$ a probability distribution for the disordered parameters $\{h_i\}$.
The third line in Eq.~\eqref{eq_pre_occ} is obtained by the use of the self-averaging property \cite{Ohkubo2006}.

In order to calculate $\langle \ln Z_{L,N} \rangle $ and $\langle \ln Z_{L,N}' \rangle$, we use the replica technique.
The following replica identity is used in order to calculate the configurational average:
\begin{align}
\left\langle \ln Z_{L,N} \right\rangle_{\{ h_2, \dots, h_L\}}
=& \lim_{m\to 0} \left( \frac{\left\langle Z_{L,N}^m \right\rangle_{\{ h_2, \dots, h_L\}} - 1}{m}
\right) 
\label{eq_replica}.
\end{align}

Using the integral representation of Kronecker delta,
$2 \pi i \delta(a,b) = \oint \textrm{d}z \, z^{a-b-1}$,
and defining 
\begin{align}
H(h,z) =  \sum_{n=0}^\infty \frac{p(h,n)}{n!} z^n
\label{eq_generating_function},
\end{align}
the replicated partition function in the configurational average is given by
\begin{align}
&\left\langle Z_{L,N}^m \right\rangle_{\{ h_2, \dots, h_L\}} \notag \\
=& \oint \frac{dz_1}{2\pi i} \cdots \oint \frac{dz_m}{2\pi i} 
e^{- (N+1) \sum_{\alpha=1}^m \ln z_\alpha + \sum_{\alpha=1}^m \ln H(h_1,z_\alpha)} \notag \\
& \times \exp\left(
(L-1) \ln \left\langle \exp\left[\sum_{\alpha=1}^m  \ln H(h,z_\alpha) \right] \right\rangle_{h} 
\right).
\end{align}
Assuming the replica symmetry ($z_\alpha = z$) and using the saddle-point method, 
we get the following saddle-point equation
\begin{align}
\rho = \left\langle 
z_\textrm{s} \left( \frac{d}{dz_\textrm{s}} H(h,z_\textrm{s}) \right)
\Big/ H(h,z_\textrm{s})
\right\rangle_h,
\label{eq_saddle_point}
\end{align}
and 
\begin{align}
\left\langle \ln Z_{L,N} \right\rangle_{\{ h_2, \dots, h_L\}} 
\simeq &- (\rho L + 1) \ln z_\textrm{s} + \ln H(h_1, z_\textrm{s}) \notag \\
& + (L-1) \langle \ln H (h,z_\textrm{s})\rangle_h + C,
\end{align}
where $C$ is a constant.
We perform the similar calculation for $\langle \ln Z_{L,M}' \rangle$,
and obtain the same saddle-point equation as Eq.~\eqref{eq_saddle_point}
and 
\begin{align}
\left\langle \ln Z_{L,N}' \right\rangle_{\{ h_2, \dots, h_L\}} 
\simeq &- (\rho L + 1 - k) \ln z_\textrm{s} \notag \\
& + (L-1) \langle \ln H (h,z_\textrm{s})\rangle_h + C,
\end{align}
where the constants $C$ in $\langle Z_{L,N} \rangle_{\{ h_2, \dots, h_L\}}$
and $\langle Z_{L,N}' \rangle_{\{ h_2, \dots, h_L\}}$  are the same.
Finally, we get the following occupation distribution
\begin{align}
P(k) 
= \int dh \,\, 
\phi(h)
\frac{p(h,k)}{k!}
\frac{z_\textrm{s}^k}{H(h,z_\textrm{s})}
\label{eq_occupation_distribution}.
\end{align}

For the condensation phenomenon,
the radius of convergence $z$ in Eq.~\eqref{eq_generating_function} plays an important role.
We here define the radius of convergence as $z = \beta$.
In Eq.~\eqref{eq_saddle_point},
increasing the value of $z_\textrm{s}$ corresponds to the increase of the density $\rho(z_\textrm{s})$.
The radius of convergence $z_\textrm{s} = \beta$ gives the maximal value of the density $\rho_\textrm{c}$:
\begin{align}
\rho_\textrm{c} = \beta \left\langle \left( \frac{d}{d\beta} H(h,\beta) \right)
\Big/ H(h,\beta)
\right\rangle_h.
\end{align}
When the density $\rho$ goes beyond the critical density $\rho_\textrm{c}$,
the excess particles condense only on a site.

It has been revealed that the following conditions are needed for the condensation phenomenon \cite{Evans2005}:
\begin{itemize}
\item If the radius of convergence $\beta$ is infinite, the condensation phenomena do not occur.
\item If the radius of convergence $\beta$ is finite and $\rho_\textrm{c}$ is infinite, 
one can find a solution of Eq.~\eqref{eq_saddle_point} for any finite density, 
so that there is no condensation phenomena.
\item If both $\beta$ and $\rho_\textrm{c}$ are finite, 
one can no longer satisfy Eq.~\eqref{eq_saddle_point} for $\rho > \rho_\textrm{c}$.
In this case, we have condensation phenomena.
\end{itemize}
It is obvious that the condensation phenomenon is difficult to occur in the Ehrenfest class,
because the factor $1/n!$ makes the summation of Eq.~\eqref{eq_generating_function} converge  
for much choice of $p(h,n)$.
Hence, the choice of the transition probability (or $p(h,n)$) would be important for the condensation phenomenon.

\section{Specific example: Disordered Preferential Urn model}

\subsection{Definition of the preferential urn model}

In spite of the difficulty of occurrence of the condensation phenomenon in the Ehrenfest class with quenched disorder,
it is possible to construct a model for the condensation phenomenon.
In order to research the condensation phenomenon in more detail,
we consider a more concrete example, i.e., the disordered preferential urn model on a complete graph
proposed in Ref.~\cite{Ohkubo2005}.
The weight in the partition function is defined by
\begin{align}
p(h_i,n_i) = (n_i!)^{h_i}.
\label{eq_p}
\end{align}
For this choice of $p(h,n)$,
the condensation phenomenon can arise  because 
\begin{align}
H(h,z) = \sum_{n=0}^\infty (n!)^{h-1} z^n
\label{eq_H}
\end{align}
diverges when $h=1$ and $z \geq 1$.

The above choice of $p(h,n)$ means the transition probability $u'(h_i,n_i) = (n_i+1)^{h_i}$.
The above dynamics may not be adequate for physical systems, 
but this transition probability is related to the concept of preference, i.e., ``the rich get richer'',
which is important in the research field of complex networks \cite{Barabasi1999}. 
Hence, it would be worth to consider such choice of $p(h,n)$.

\subsection{Non-disordered case}

We here study the case with no disorder: All disorder parameters are $h_i = h = 1$.
Since the function $H(h,z)$ of Eq.~\eqref{eq_H} diverges with $h = 1$ and $z \geq 1$,
we have to check the critical density.
When $h = 1$, the saddle point equation \eqref{eq_saddle_point} gives
\begin{align}
\rho = \frac{z_\textrm{s}}{1-z_\textrm{s}}.
\end{align}
This means that the critical density $\rho_\textrm{c}$ in infinite.
Hence, there is no condensation phenomenon, as described in Sec.III.

It is easy to confirm that the condensation does not occur in the non-disordered case with $h \leq 1$.
When $h > 1$, a type of condensation could arise; we discuss the case in the following subsection.

\subsection{Disordered cases: Three types of condensation}

It has been revealed that the sites with the disorder parameter $h_i = 1$ are important for the condensation 
phenomenon in the preferential urn model,
but the non-disordered case is not sufficient for the occurrence of the condensation.
Henceforth, we analytically and numerically discuss three disordered cases
which corresponds to different types of condensation phenomena, respectively.

\subsubsection{Complete condensation}

When there is a site $h_i > 1$,
Eq.~\eqref{eq_H} has the radius of convergence $\beta = 0$,
and the critical density $\rho_\textrm{c}$ becomes zero from Eq.~\eqref{eq_saddle_point}.
This condensation phenomenon shall be called the `complete condensation'.
In the complete condensation, only one site contains almost of all particles for any $\rho$,
and the ratio of the number of particles containing the condensed site
tends to be one in the limit of a large system.

\subsubsection{Genuine condensation induced by the disorder}

We here consider the disordered case in which each disorder parameter $h$ is assigned by the probability density
\begin{align}
\phi(h) = (1+\alpha) (1-h)^\alpha \,\, (0 \leq h \leq 1)
\label{eq_phi}.
\end{align}

When $\alpha = 0$, i.e., we set $\phi(h)$ as an uniform distribution,
we have no condensation phenomenon;
the saddle point $z_\textrm{s}$ gradually increases as $\rho \to \infty$,
but it converges to $z_\textrm{s} = 1$,
so that the critical density is $\rho_\textrm{c} = \infty$.

Next, we check the case with $\alpha = 2$ \cite{note_alpha}.
In this disordered case,
we get the critical density $\rho_\textrm{c} = 1.26$,
so that we have the condensation phenomenon in this disordered case.
We shall call the condensation as `genuine condensation',
and distinguish it from the complete condensation and pseudo-condensation explained later.

The occupation distribution in the steady state is obtained by Eq.~\eqref{eq_occupation_distribution}.
When we restrict the value of $\alpha$ in Eq.~\eqref{eq_phi} to an integer value and $\alpha \geq 0$,
we can calculate the approximate form of the occupation distribution as
\begin{align}
P(k) \sim k^{-\alpha-2} (\ln k)^{-\alpha-2}
\label{eq_power_law}
\end{align}
for large $k$ region,
by using Eqs.~\eqref{eq_occupation_distribution}, \eqref{eq_p}, \eqref{eq_phi} and
setting $z_\textrm{s} = 1$.
We note that the occupation distribution obtained analytically can not represent a condensed site;
the occupation distribution is adequate except for the condensed site.
When we have the condensation phenomenon,
only a site deviates from the occupation distribution obtained analytically.

We here confirm the condensation phenomenon using numerical experiments.
We have performed numerical experiments for the disordered urn model on a complete graph
with lattice size $L=4000$.
All results are obtained by averaging $20$ realizations.
It is notable that the condensation arises only in the case 
in which there is at least one site with $h_i = 1$, as discussed in Sec.~III.
Although we should assign the disorder parameter $h_i$ to each site
by using the probability density $\phi(h)$ defined as Eq.~\eqref{eq_phi},
a site with the disorder parameter $h_i = 1$ is difficult to emerge 
because we here treat a finite-size system.
The value of $h_i$ near $1$ is not sufficient to occur the condensation phenomenon, as discussed later.
We therefore perform the following rescaling for the disorder parameters:
\begin{enumerate}
\item Assign $h_i$ to each site by $\phi(h)$ of Eq.~\eqref{eq_phi}.
\item Search the maximum value of $h_i$ and set the value as $h_\textrm{max}$.
\item For all $i$, we perform the rescaling $h_i \to h_i / h_\textrm{max}$.
\end{enumerate}
When we have the maximum value of $h_\textrm{max} \simeq 1$,
it is expected that the rescaling procedure has little influence on the disorder parameters $\{h_i\}$.

\begin{figure}
\begin{center}
  \includegraphics[width=70mm,keepaspectratio,clip]{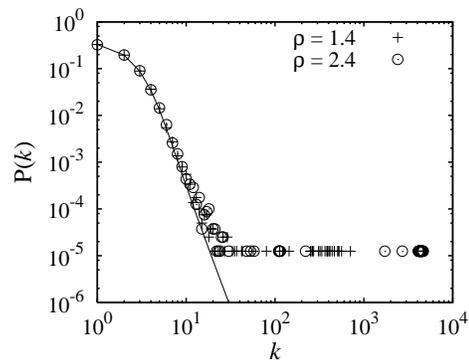} 
\caption{
The occupation distribution for the disordered case with $\alpha = 2$.
The solid line corresponds to the analytical result in Eq.~\eqref{eq_occupation_distribution} with $z_\textrm{s} = 1$.
For details, see text.
}
\label{fig_distribution}
\end{center}
\end{figure}

Figure~\ref{fig_distribution} shows the results for $\alpha = 2.0$.
We have calculated the cases with the density $\rho = 1.4$ and $\rho = 2.4$.
The solid line in Fig.~\ref{fig_distribution} corresponds to the analytical result of 
Eq.~\eqref{eq_occupation_distribution} for $z_\textrm{s} = 1$.
The occupation distribution shows the power law behavior in a certain $k$ region, 
which is expected from Eq.~\eqref{eq_power_law}.
Note that because we performed the numerical experiments with $L = 4000$ and the results are averaged out over
$20$ simulations,
the minimum value of $P(k)$ is $1 / (4000 \times 20) = 1.25 \times 10^{-5}$. 

In Fig.~\ref{fig_distribution}, 
we can see that there are several points which deviate from the analytical result (solid line)
because we take the average over $20$ simulations.
We here note that in each simulation, there is only one point which deviates from the analytical result.
The deviating point means the fact that one site has quite a lot of particles in each simulation,
which indicates the occurrence of the condensation phenomenon.
Note that due to fluctuation among each simulation,
the largest value of $k$ in each simulation is different from each other,
and hence we do not have a peak but the flat tail.

We also see the condensation phenomenon by comparing with the case of $\rho = 1.4$ and $\rho = 2.4$; 
the maximum value of $k$ increases in the case of $\rho = 2.4$ compared with that of $\rho = 1.4$
in Fig.~\ref{fig_distribution}.
However, the shapes of the distribution in small $k$ region are the same in both cases.
The analytical treatment suggests that 
the occupation distribution does not vary except for the condensed site in the condensed phase.
Hence, we confirm the occurrence of the condensation 
from the unchangeability of the occupation distribution in small $k$ region
and the shift to the right of the deviating points in the large $k$ region.

\begin{figure}
\begin{center}
  \includegraphics[width=70mm,keepaspectratio,clip]{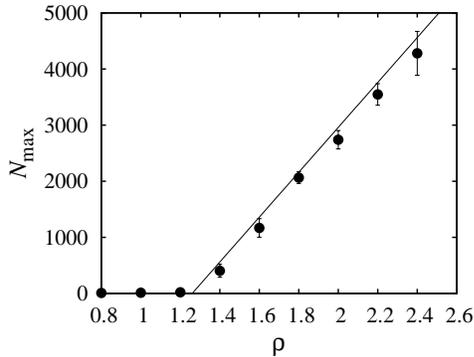} 
\caption{
The maximum number of particles in the site with condensation.
The solid line represents Eq.~\eqref{eq_maximum}.
}
\label{fig_condensation}
\end{center}
\end{figure}

In order to see the condensation phenomenon more clearly,
we have calculated the number of particles in the site which contains the maximum number of particles.
The maximum number of particles in the condensed site is expected to be
\begin{align}
N_\textrm{max} = L (\rho - \rho_\textrm{c}),
\label{eq_maximum}
\end{align}
because the system can contain only up to $L \rho_\textrm{c}$ particles without condensation.
Figure~\ref{fig_condensation} shows the numerical results for $\alpha = 2$ and 
the analytical result of Eq.~\eqref{eq_maximum} (the solid line).
The maximum number $N_\textrm{max}$ shows a linear dependence on the density $\rho$.
This result confirms that the macroscopic number ($\sim \mathcal{O}(L)$) of particles is condensed 
only in a single site \cite{note_numerical}.

\subsubsection{Pseudo-condensation}

\begin{figure}
\begin{center}
  \includegraphics[width=70mm,keepaspectratio,clip]{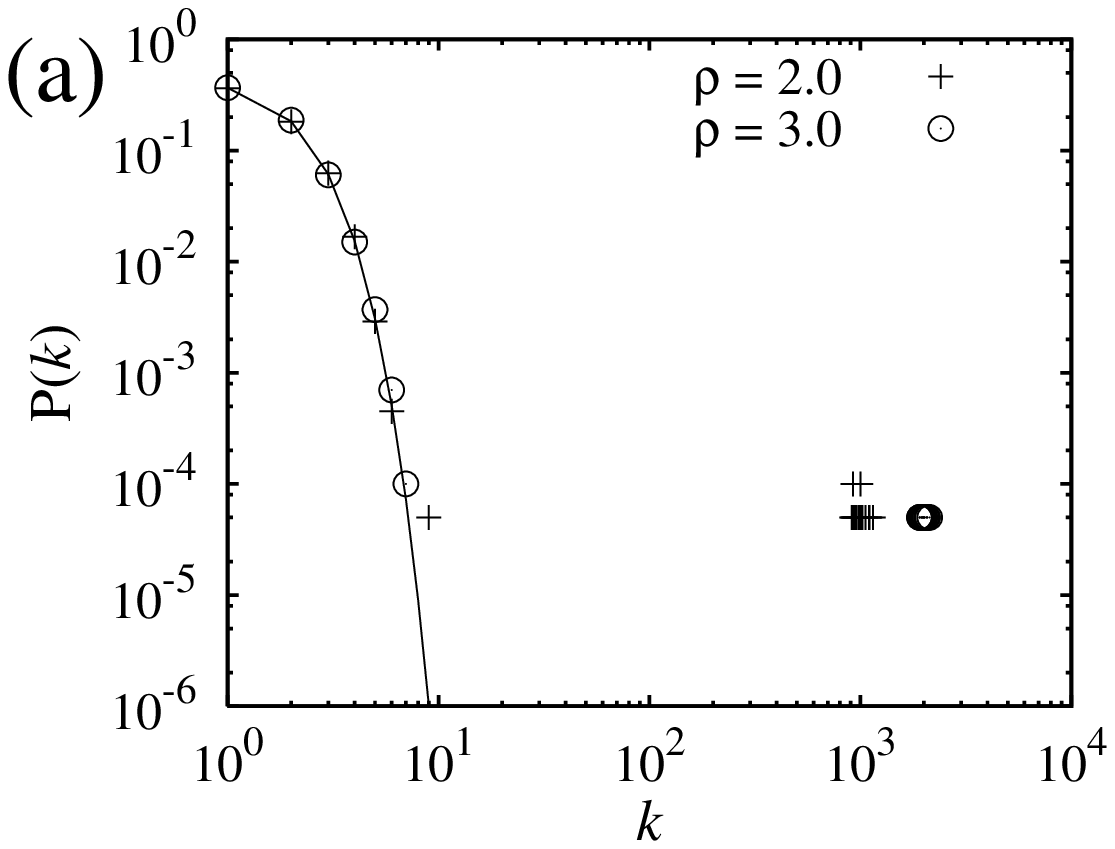} 
  \includegraphics[width=70mm,keepaspectratio,clip]{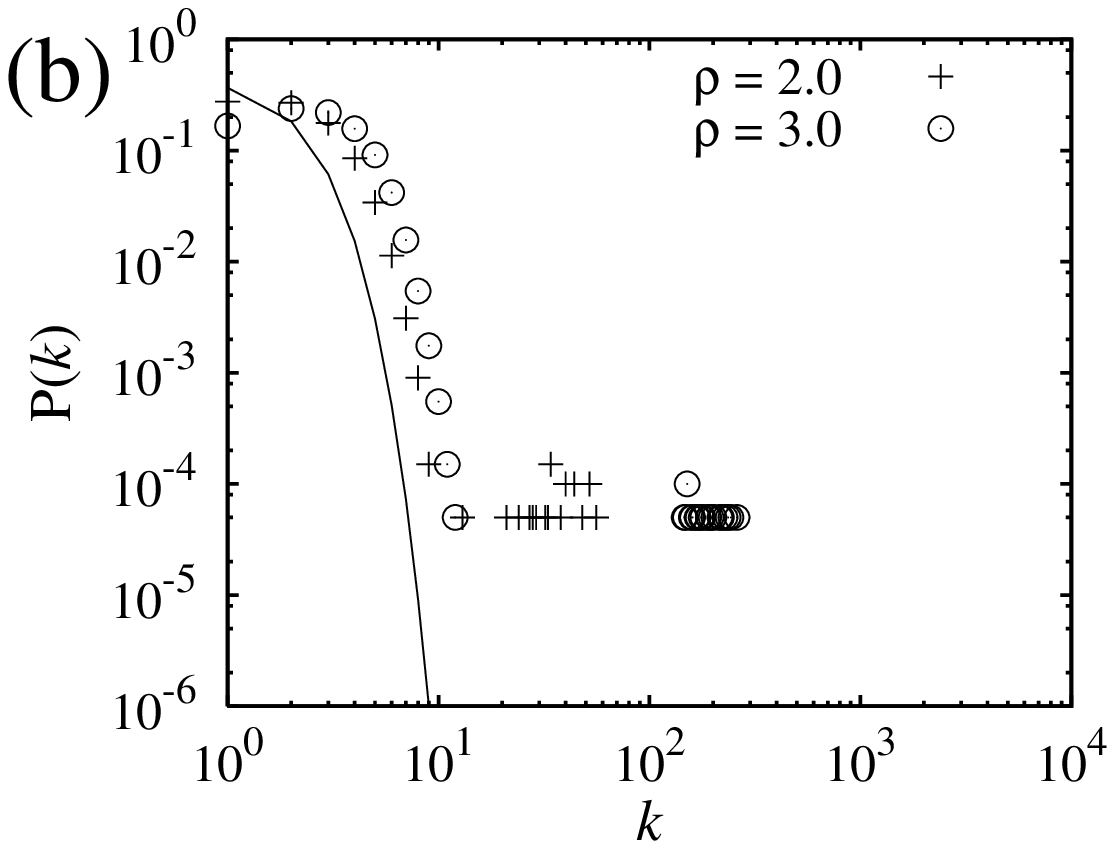} 
\caption{
The occupation distributions for the cases with single-site disorder.
(a)The genuine condensation phenomenon ($h_\textrm{disorder} = 1.0$). 
(b)Pseudo-condensation phenomenon ($h_\textrm{disorder} = 0.8$).
The solid line corresponds to Eq.~\eqref{eq_poisson}.
All data are averaged over $20$ realizations.
}
\label{fig_condensation2}
\end{center}
\end{figure}

We here confirm that a site with the disorder parameter $h_i=1$ is necessary for the condensation phenomenon,
as discussed in the previous subsection.
We consider a single-site disordered case 
in which all disorder parameters $h_i = 0$ except for only one site with $h_\textrm{disorder} = 1$.
In this case, one can easily confirm that
the critical density is equal to one ($\rho_\textrm{c} = 1$) from Eq.~\eqref{eq_saddle_point}.
The occupation distribution of Eq.~\eqref{eq_occupation_distribution}
becomes the Poisson distribution
\begin{align}
P(k) = e^{-\rho_\textrm{c}} \frac{\rho_\textrm{c}^k}{k!}
\label{eq_poisson}
\end{align}
when the condensation phenomenon arises.
We have performed numerical experiments with the lattice site $L = 1000$.

Figure~\ref{fig_condensation2}(a) shows the case in which only a single site has $h_\textrm{disorder} = 1$.
The numerical results are obtained by taking the average over $20$ simulations.
The solid line corresponds to the Poisson distribution of Eq.~\eqref{eq_poisson}.
One can easily see the occurrence of the condensation phenomenon, as explained in Sec.IV.C.2:
compared with the case with $\rho = 2.0$,
the maximum value of the occupation number increases in the case with $\rho = 3.0$.
In addition, the occupation distributions are in good agreement with the analytical results,
except for the condensed site.
This condensation phenomenon corresponds to the genuine condensation, as described in Sec.IV.C.2.

In contrast, the genuine condensation phenomenon does not occur in the case 
where only a single site has $h_\textrm{disorder} = 0.8$,
as shown in Fig.~\ref{fig_condensation2}(b).
One may see that the genuine condensation phenomenon occurs in Fig.~\ref{fig_condensation2}(b),
because there are largely deviating points.
However, in this case,
the whole shape of the occupation distribution in small $k$ region changes with the increase of the density $\rho$.
When the genuine condensation arises,
the increase of the density $\rho$ does not change the occupation distribution in small $k$ region,
as explained in the previous subsection.
The change of the occupation distribution means that
we do not have the genuine condensation phenomenon in spite of the existence of the deviating points 
in Fig.~\ref{fig_condensation2}(b).
Evans \textit{et al.} \cite{Evans2006b} have called the similar type of condensation as `pseudo-condensation'.
In this case, the partition function is analytic even if the number of particles increases,
so that there in no true phase transition.
In summary, when there are deviating points in the occupation distribution and the whole shape of 
the occupation distribution varies with the increase of the density $\rho$,
we call the phenomenon as the `pseudo-condensation'.
It was revealed that the pseudo-condensation can occur in the preferential urn model
depending on the nature of the quenched disorder.

\section{Concluding remarks}

In summary, 
the condensation phenomenon in `Maxwell-Boltzmann' statistics has been studied.
We especially focused on disordered cases in the Ehrenfest class.
We have developed the analytical treatment for the disordered cases,
and clarified the conditions for the condensation.
In addition, using the preferential urn model, we have demonstrated the condensation phenomena.
It was revealed that only the disordered cases give the genuine condensation phenomenon in the preferential urn model.
Furthermore, depending on the types of the quenched disorder, we have three types of condensation.

In order to analyze the condensation phenomenon in disordered cases in more detail,
we might need other analytical methods.
For example, 
the canonical analysis proposed by Evans \textit{et al.}\cite{Evans2006b} 
is useful for the study of the condensation in a finite size system;
one can analyze the condensation transition and the structure of the condensate, 
determining the precise shape and the size of the condensate in the condensed phase
by the use of the canonical analysis.
Although the grand canonical analysis for disordered cases has been developed
in Ref.~\cite{Leuzzi2002},
we have not yet obtained an adequate analytical method for the disordered cases
in the canonical ensemble.
The disordered version of the canonical analysis would be important for future research.

\section*{ACKNOWLEDGMENTS}
This work was supported in part by grant-in-aid for scientific research (No. 18$\cdot$5140)
from the Ministry of Education, Culture, Sports, Science and Technology, Japan.


\begin{thebibliography}{99}
\bibitem{note_alpha} 
It is difficult for us to express the $\rho_\textrm{c}$ as a function of $\alpha$ 
by the use of elementary functions.
Hence, we evaluate the critical density by solving Eq.~\eqref{eq_saddle_point} numerically.
It is an open question that how large $\alpha$ gives the condensation phenomenon,
but we have numerically checked that $\alpha \geq 1$ has a finite critical density.
\bibitem{note_numerical} 
We consider that the quantitative difference between the numerical results and the analytical result
in Fig.~\ref{fig_condensation} stems from the finite size effect and the rescaling procedure.
\end{thebibliography}
\end{document}